\documentclass[aps,preprint]{revtex4}
\usepackage{graphics,epsfig,amssymb}

\def\beq{\begin{equation}} 
\def\eeq{\end{equation}} 
\begin{document}

\title{Band-like structures and quartets in deformed $N=Z$ nuclei}

\author{M. Sambataro$^a$ and N. Sandulescu$^b$}
\affiliation{$^a$Istituto Nazionale di Fisica Nucleare - Sezione di Catania,
Via S. Sofia 64, I-95123 Catania, Italy \\
$^b$National Institute of Physics and Nuclear Engineering, P.O. Box MG-6, 
Magurele, Bucharest, Romania}

\begin{abstract}
We provide a description of deformed $N=Z$ nuclei in a formalism of $\alpha$-like quartets. Quartets are constructed variationally by resorting to the use of proper intrinsic states. Various types of intrinsic states are introduced which generate different sets of quartets for a given nucleus. Energy spectra are generated via configuration-iteraction calculations in the spaces built with these quartets. The approach has been applied to $^{24}$Mg and $^{28}$Si in the $sd$ shell and to $^{48}$Cr in the $pf$ shell. In all cases a good description of the low-lying spectra has been achieved. As a peculiarity of the approach, a close correspondence is observed between the various sets of quartets employed and the occurrence of well defined band-like structures in the spectra of the systems under study.
\end{abstract}

\maketitle

\section{Introduction}

The study of $N=Z$ nuclei, i.e., nuclei with an equal number of neutrons and protons, is one of the key issues of contemporary nuclear structure physics.  What makes $N=Z$ nuclei particularly interesting
is the appearance of new types of correlations induced by the proton-neutron ($pn$) pairing interaction. This interaction is expected to play a relevant role in these nuclei owing to the fact that protons and neutrons share the same orbitals. The $pn$ pairing interaction can manifest itself both in an isovector ($T=1$) and in an isoscalar ($T=0$) mode and the study of the corresponding correlations is matter of active investigation (e.g., see \cite{goodman2001,frauendorf,sagawa}). 

Understanding the type of correlations induced by the $pn$ pairing force in the wave function of $N=Z$ nuclei has proved to be a not trivial task. Following the seminal work by Belyaev et al. \cite{belyaev}, in most studies the ground state of a $pn$ pairing Hamiltonian has been described in the Hartree-Fock-Bogoliubov (HFB) approximation.  Since this approximation does not conserve the particle number, the spin and the isospin, a lot of efforts is being  made to restore these symmetries
through projection techniques (e.g., see \cite{romero,proj_hfb}). In this way additional  correlations 
can be taken into account which are not included in the standard two-body pairing correlations associated with the HFB approach. However, realistic calculations on this line are still missing and, in addition, it is not yet clear how one could identify what kind of correlations are contained in the projected-HFB state.

An unambigous indication of the type of correlations that are generated 
by the $pn$ pairing in $N=Z$ nuclei is provided by the exact analytical solutions of realistic Hamiltonians.
In the case of the isovector pairing, finding such solutions has gone through various works spread over several decades \cite{richardson,pan,links,dukelsky,skrypnyk1,skrypnyk2}, the first complete description of both eigenvalues and eigenstates being provided in 2006 by Dukelsky et al. \cite{dukelsky}. However, the complicated  mathematical structure of these exact solutions has hindered a  simple understanding of the underlying correlations.

In a recent work \cite{sasa_so5_exact}, we have provided an alternative and more transparent description 
of the exact solutions of the isovector pairing which has evidenced a peculiar aspect of these solutions which had escaped the previous investigations. Indeed, we have shown on an analytic basis that what characterizes the isovector pairing Hamiltonian in $N=Z$ systems is the occurrence in its eigenstates of correlated four-body $\alpha$-like structures (``quartets''). 
More precisely, in Ref. \cite{sasa_so5_exact}, we have shown that exact isospin $T=0$ seniority-zero eigenstates of a constant strength isovector Hamiltonian, for an even number of protons and neutrons distributed over a set of non-degenerate levels, can be expressed as linear superpositions of products of $T=0$ quartets, each quartet being a product of two $T=1$ collective pairs. 

One needs to remark that, well before the exact treatment just discussed the essential role played by four-body correlations in $N=Z$ systems subject to an isovector pairing force, but limited to the special case of degenerate single-particle levels (the so-called SO(5) model), had already been emphasized in Ref. \cite{dobes}. More generally, quartets have definitely a long history in nuclear structure physics 
\cite{soloviev,flowers,arima,farraggi,eichler,catara,hasegawa,senkov},
but their complexity has undoubtedly represented a hindrance to the development of quartet models.

In a recent past, $T=0$ quartets have been introduced to build an approximation scheme for the ground state of the isovector pairing Hamiltonian \cite{qcm_t1}. This approach, known as Quartet Condensation Model (QCM), assumes this ground state to be a condensate of $T=0$ quartets, each quartet consisting of two collective isovector pairs coupled to $T=0$. The QCM approach  has turned out to be very accurate both in the case of deformed and spherical mean fields. In the latter case the quartets are also characterized by an angular momentum $J=0$ \cite{sasa_t1}. 
The QCM formalism has been later on extended to a more general form of $pn$ pairing which also included an isoscalar component \cite{qcm_t0t1,qm_t1t0,qm_qcm_t0t1} as well as to quite general Hamiltonians \cite{qcm_sm}. 
The link between the QCM condensate and the exact ground state has been recently analyzed in 
the case of the isovector pairing \cite{qcm_exact}.

The use of quartets has not been limited to the analysis of the ground state only. A more elaborate quartet formalism, involving quartets other than the single $J=0,T=0$ one of the QCM approach, has been developed to describe the spectra of $N=Z$ systems \cite{qm_prl,qm_pd,qm_odd}. In this approach, simply referred to as Quartet Model (QM), spectra have been generated by carrying out configuration-interaction calculations in a space of states formulated as products of collective $T=0$ quartets of various $J$. 

A crucial aspect of the QM approach consists in the definition of the quartets to involve in the calculations. In all the cases studied so far we have adopted the criterion of assuming as $T=0$ quartets those defining the low-lying eigenstates of the nearest $T=0$ one-quartet systems
\cite{qm_prl,qm_pd,qm_odd}.
While having the advantage of being straightforward, this ``static'' definition of the quartets is clearly not the most appropriate one since it fully neglects the effect of the Pauli principle on the amplitudes of the quartets when two or more of these quartets have to coexist in the same nucleus. Finding the most appropriate quartets to be employed in a QM calculation is a matter of primary importance for the validity of this approach and this has represented one of the goals of the present work.

The problem of the selection and definition of the quartets that we are facing within the QM is fully analogous to that encountered in the selection of the most appropriate collective pairs in the context of the Nucleon Pair Approximation (NPA) \cite{npa}. 
This is a long-lasting \cite{klein,zirnbauer,maglione,maglione2,sam,allaart} and still topical problem \cite{lei,npa_calvin,npa_calvin2} in nuclear structure physics. In the present paper we will follow an approach inspired to that of Refs. \cite{maglione,maglione2,catara2}  but applied in a context of quartets
rather then pairs. According to that, quartets will result from the minimization of special intrinsic states. We will present three different applications of our approach, two referring to the $sd$ shell ($^{24}$Mg and $^{28}$Si) and one to the $pf$ shell ($^{48}$Cr). We will show that the new criterion to fix the quartets will provide not only a good description of the low-lying states of these nuclei but, as a peculiarity, it will also allow the identification of band-like structures in their spectra.

In Section II, we will describe our approach and illustrate the applications. In Section III,
we will summarize the results and draw some conclusions.

\section{Procedure and results}

We work in a spherically symmetric mean field and, using the standard notation, we introduce the label
$i\equiv \{n_i,l_i,j_i\}$ to identify the orbital quantum numbers. We define the  $T=0$ quartet creation operator as
\begin{equation}
q^+_{JM}=\sum_{i_1j_1J_1}\sum_{i_2j_2J_2}\sum_{T'}
q_{i_1j_1J_1,i_2j_2J_2,{T'}}
[[a^+_{i_1}a^+_{j_1}]^{J_1{T'}}[a^+_{i_2}a^+_{j_2}]^{J_2{T'}}]^{JT=0}_{M},
\label{1}
\end{equation}
where $a^+_i$ creates a fermion on the orbital $i$ and $M$ stands for the projection of $J$.
No restrictions on the intermediate couplings $J_1T'$ and $J_2T'$ are introduced and the amplitudes 
$q_{i_1j_1J_1,i_2j_2J_2,{T'}}$ are supposed to guarantee the normalization of the operator.
We shall focus on systems of $N_\pi$ protons and $N_\nu$ neutrons such that $N_\pi=N_\nu$
and  $N_\pi+N_\nu =4n$ ($n=2,3$) and assume axially symmetry of these systems.
Once quartets have been fixed, we generate the energy spectra by carrying out configuration-interaction calculations.
To this purpose we define the set of states (we work in the $m$-scheme)
\begin{equation}
|\Psi^{(n)}_{\overline M},\{N_{JM}\}\rangle = \prod_{J\in{(0,J_{max})}; M\in{(-J,J)} }(q^+_{JM})^{N_{JM}}|0\rangle
\label{1a}
\end{equation}
with the conditions
\beq
\sum_{JM}N_{JM}=n,~~~~~~~\sum_{JM}MN_{JM}=\overline{M}.
\eeq
We then orthonormalize the states (\ref{1a}) and diagonalize the Hamiltonian in this new basis for the various $\overline{M}$.
Calculations have been carried out in the $sd$ and $pf$ shells by adopting the USDB \cite{usdb} and KB3G \cite{kb3g} interactions, respectively.

As a starting point we introduce the state 
\beq
|\Theta_{g}\rangle = \mathcal{N}_g(Q^+_g)^n|0\rangle ,
\label{2}
\eeq
where
\beq
Q^+_g =\sum_J \alpha_{g,J}(q^+_g)_{J0}
\label{3}
\eeq
and $\mathcal{N}_g$ is a normalization factor. $|\Theta_{g}\rangle$ is thus a condensate of the quartet $Q^+_g$ which is a linear superposition of quartets $(q^+_g)_{J0}$ of the general form (\ref{1}) and whose angular momentum $J$ runs over a set of values to be specified. After fixing this set of values, we minimize the energy of the state $|\Theta_{g}\rangle$ and so define the quartets $(q^+_g)_{J}$
(let's call these ``dynamical'' quartets). We are then ready to carry out configuration-interaction calculations in a space spanned by these quartets. In Figs. 1,2 and 3, columns (A) and (B), we show what changes occur in the low-lying spectra of the nuclei under study when a set of dynamical quartets replace the analogous set of static quartets. These static quartets are those describing the lowest states of $^{20}$Ne for the $sd$ shell nuclei and those of $^{44}$Ti for $^{48}$Cr. In the columns (A) we plot all the positive parity states up to the $6_1$ obtained using a set of static $J$=0,2,4 (plus $J$=6 in the case of $^{48}$Cr) quartets. In spite of the quite good results for the ground state correlation energy (defined as the difference between the ground state energies with and in absence of interaction), in all cases the spectra show marked differences with respect to the shell model results. In the columns (B) of the same figures, the dynamical quartets with the same $J$'s as above have been employed. In all three nuclei one observes a lowering of the yrast states $J=0,2,4,6$ forming the ground state bands while most of the remaining states are pushed up in energy. The new ground state bands are all closer in energy to the shell model ones and a considerable improvement is observed also in the accuracy of the ground state correlation energies. Thus adopting the quartets $(q^+_g)_{J0}$ associated with 
$|\Theta_{g}\rangle$ has had a positive effect only on the ground state bands of the nuclei under study. The state (\ref{2}) will be referred to as ``ground'' intrinsic state.

In Fig. 4 we plot the amplitudes $\alpha_{g,J}$ (normalized such that $\sum_J\alpha^2_{g,J}=1)$ which characterize the ground intrinsic quartets $Q^\dag_g$ in the three cases. In all cases $\alpha_{g,J=0}$
turns out to be the largest amplitude. This is especially the case for  $^{28}$Si, which clearly stands
out for a much more prominent role of this amplitude as compared with the $J>0$ ones. We shall further 
comment on this point below.

The not yet satisfactory agreement between the spectra (B) of Figs. 1, 2 and 3 and the corresponding shell model spectra makes clear that new quartets must come into play. To this end, by starting from $|\Theta_{g}\rangle$ and promoting one of the quartets $Q^+_g$ to an excited configuration, we  generate new intrinsic states.
We define a ``$\beta$'' intrinsic state as
\beq
|\Theta_{\beta}\rangle = \mathcal{N}_\beta Q^\dag_\beta (Q^\dag_g)^{(n-1)}|0\rangle ,
\label{4}
\eeq
where
\beq
Q^\dag_\beta =\sum_J \alpha_{\beta ,J}(q^\dag_\beta)_{J0}
\label{5}
\eeq
and $(q^\dag_\beta)_{J0}$  is a quartet of the general form (\ref{1}).
Assuming that the quartets $(q^+_g)_{J0}$ have already been fixed, we construct the new quartets $(q^\dag_\beta)_{J0}$ by minimizing the energy of $|\Theta_{\beta}\rangle$ under the constraint of orthogonality to 
$|\Theta_g\rangle$.

Similarly, we define a ``$\gamma$'' intrinsic state as
\beq
|\Theta_{\gamma}\rangle = \mathcal{N}_\gamma Q^\dag_\gamma (Q^\dag_g)^{(n-1)}|0\rangle ,
\label{6}
\eeq
where
\beq
Q^\dag_\gamma =\sum_J \alpha_{\gamma ,J}(q^\dag_\gamma)_{J2}
\label{7}
\eeq
and $(q^\dag_\gamma)_{J2}$ is a quartet still of the general form (\ref{1})
but characterized by an angular momentum projection $M=2$.
More generally, we define a ``$k$'' ($k$=3,4,...) intrinsic state as
\beq
|\Theta_k\rangle = \mathcal{N}_k Q^\dag_k (Q^\dag_g)^{(n-1)}|0\rangle ,
\label{8}
\eeq
with
\beq
Q^\dag_k =\sum_J \alpha_{k,J}(q^\dag_k)_{Jk},
\label{9}
\eeq
where the quartets $(q^\dag_k)_{Jk}$, still of the general form (\ref{1}), are characterized by an angular momentum projection $M=k$. In these two last cases the new intrinsic quartets $Q^\dag_\gamma$ and $Q^\dag_k$ are constructed variationally without any special constraint.

We proceed by observing, case-by-case, how the use of these new intrinsic states can help  to improve the spectra of the nuclei under study.
In $^{24}$Mg, Fig. 1(B), one notices that a band formed by the $J_k=2_2,3_1,4_2,5_1$ states has been shifted higher in energy when passing from static to dynamical quartets. By making use of the definition of the
$\gamma$ intrinsic state (\ref{6}), we construct new quartets 
$(q^\dag_\gamma)_{J}$ with $J=2,3,4$ and perform a configuration-interaction calculation that includes
these new quartets in addition to those already used for the calculation of Fig. 1(B). The new result is shown in Fig. 1(C). We observe a clear lowering of the energies of the states $J_k=2_2,3_1,4_2,5_1$ while the states of the ground state band have remained basically unmodified with respect to those of column (B). The inclusion of the quartets derived with the help of the $\gamma$ intrinsic state (\ref{6}) has therefore essentially affected only those states which can be associated to a $\gamma$ band of $^{24}$Mg (and, in addition, the state $2_3$). As a final step, we have explored the effect on this spectrum of the inclusion of a set of quartets 
$(q^\dag_\beta)_{J}$ with $J=0,2,4$. The basic effect which can be observed in Fig. 1(D) is a lowering the states $J=0_2,2_3$ which can be associated to a $\beta$ band of $^{24}$Mg. As a result of the new diagonalization, also the $5_1$ state is lowered in energy. The final spectrum shows a good agreement with the shell model one.

For what concerns $^{28}$Si, what is most striking in the spectrum of Fig. 2(B) is the absence of a state $0_2$ close to the state $4_1$ as observed in both the shell model and the experimental spectrum. 
By interpreting this state $0_2$ as the head of a $\beta$ band, as a next step, we enlarge the model space by also including the quartets $(q^\dag_\beta)_{J}$ with $J=0,2,4$
associated with the $\beta$ intrinsic state. The result of the new configuration calculation can be seen in Fig.2(C). The inclusion of the new quartets mostly affects the yrare $J=0,2$ states by giving rise, in particular, to a surprising lowering of the $0_2$ state which positions itself immediately above the $4_1$ state,
where it is expected to be. In spite of the fact that a reasonably good agreement with the shell model spectrum has already been achieved, we perform an additional calculation which also includes the quartets $(q^\dag_3)_{J3}$, with $J=3,4$, associated with the $k=3$ intrinsic state (\ref{8}). 
Such a new calculation is stimulated by an old analysis of $^{28}$Si \cite{sheline} in which the band head of a $K=3$ band is positioned immediately above the $0_2$ state. As it can be seen in Fig. 2(D), the new calculation essentially lowers the energy of the $3_1$ and $4_2$ states and, in addition, that of the $3_2$ state. This new calculation further improves the quality of the QM spectrum which compares well with the shell model one.

The case of $^{28}$Si deserves some extra comments. As already specified above, the quartets associated with each intrinsic state result from the minimization of the energy of this state (under  potential constraints). In the case of $^{28}$Si, however, two almost degenerate minima occur in the minimization
of the ground intrinsic state which have a very different nature. The lowest one, at $E=-133.627$ MeV, has the form of a true QCM condensate since only $J=0$ quartets survive in the minimization. The other minimum, only 36 KeV higher, sees instead the contribution of all the quartets involved in the calculation. As also visible in Fig. 4, the calculations discussed so far have been based on the latter minimum. An advantage of this choice is that this has provided us straightforwardly with the $J>0$ quartets needed to build the spectrum of Fig.2(B). However, also the possibility of adopting the lowest minimum as a starting point for our calculations has been explored. In this case we had to face the problem of how to define the $J>0$ quartets. Assuming a condensate of $J=0$ as a ground state, $J>0$ quartets can result from the ``breaking'' of one of the quartets of the condensate, namely by constructing a new state where
one of the $J=0$ quartets of the condensate is replaced with a $J>0$ one and then minimizing the energy
of this state. This technique has been successfully applied in a recent work to construct the excited states of $pn$ pairing Hamiltonians in a formalism of quartets \cite{qcm*} and is the analogous of the well known one-broken-pair approximation in the context of NPA \cite{allaart}. By adopting this technique we have therefore constructed new $J=2$ and $J=4$ quartets which, together with the $J=0$ quartet of the QCM condensate, have provided us with a set of
quartets alternative to that used for the calculations of Fig. 2(B). We have verified that the spectrum generated with this different set of quartets gives quite close excitation energies of the ground state rotational band but slightly higher energies of the additional states and a worse ground state correlation energy (the error rising from the 0.4$\%$ of Fig. 2(B) to 1$\%$). Altogether, then, the choice of generating the quartets from the second minimum has revealed to be the most effective one. The occurrence of this double minimum   confirms a certain degree of ``instability'' of the ground intrinsic  state of $^{28}$Si which had already emerged in a previous semi-phenomenological study of this nucleus \cite{qmbos}. 

The last case under investigation, $^{48}$Cr, shares some analogies with the corresponding one of $^{24}$Mg. Indeed one observes in Fig. 3(B) that, also in this case, the states $2_2,3_1,4_2$ have been shifted higher in energy when replacing the static quartets with the corresponding dynamical ones associated with the ground intrinsic state. These states reminding those of a $\gamma$ band, we proceed as for $^{24}$Mg
by introducing the quartets $(q^\dag_\gamma)_{J}$,  associated with the $\gamma$ intrinsic state (\ref{6}), with $J=2,3,4$. The new calculation, Fig. 3(C), leaves unaffected the states of the ground rotational band (as for $^{24}$Mg) while it lowers significantly the states $2_2,3_1,4_2$.
The new  calculation also leads to the appearance of new states ($2_3,4_3,0_2$) in the highest part of the spectrum. These states have a correspondence with the shell model ones but with a  $0_2$ still too high in energy. By interpreting this state as a possible head of a $\beta$ band and wishing to lower its energy, we perform a final calculation which includes also the quartets  $(q^\dag_\beta)_{J}$ with $J=0,2,4$. This calculation leads to a significant lowering of the $0_2$ state (Fig. 3(D)) which improves the agreement between exact and approximate spectra.

As a final comment, we like to remark that the high quality of the ground states reached in these calculations, with errors in the correlation energies confined within the $0.4\%$ (Figs. 1,2,3) depends almost entirely on
the low-$J$ quartets of the intrinsic ground state. Indeed, by performing configuration-interaction calculations which include only the quartets $(q^+_g)_{J}$ with $J=0,2$ these errors remain confined within 0.7$\%$ for $^{24}$Mg and $^{28}$Si while, for $^{48}$Cr, one finds 1.6$\%$. In the latter case, however, it is sufficient to add the $J$=4 quartet to reduce this error to 0.6$\%$. As already evident from the comparisons between the columns (A) and (B) of the same figures, these errors increase considerably if  static quartets are used instead of the dynamical ones. 

\section{Summary and conclusions}

In this paper we have provided a description of deformed $N=Z$ nuclei in a formalism of $\alpha$-like quartets. Quartets have been constructed variationally by resorting to the use of proper intrinsic states. Spectra have been obtained by carrying out configuration-interaction calculations in spaces built with these quartets. For each nucleus more sets of quartets have been used in correspondence with the various types of intrinsic states introduced. The intriguing aspect of these calculations has been the observation of band-like structures associated with the various sets of quartets. Also intriguing has been observing that these structures interfere little with each other in the sense that once one structure has been generated by means of a set of quartets, enlarging the configuration space with a new set of quartets moves down the states associated with the new structure without, however, strongly affecting the previous structure. The procedure has been applied to
$^{24}$Mg, $^{28}$Si and $^{48}$Cr nuclei and it has provided a good description of the low-lying spectra in all three cases. A merit of this description is that of relying on only a few degrees of freedom. For what concerns the ground states, in particular, we have shown that already the $T=0$ quartets with $J=0$ and $J=2$ can guarantee
an accurate approximation of these states. 
As a general conclusion, the results achieved in this work promote the new method proposed for the definition of the quartets as well as the use of the latter as basic structures for an effective description of the ground and excited states of deformed $N=Z$ nuclei.

\vskip 0.3cm
{\it Acknowledgments}\\ 
The authors wish to thank Danilo Gambacurta for useful discussions.
This work was supported by a grant of the Romanian Ministry of Research and Innovation, CNCS - UEFISCDI, project number PCE 160/2021, within PNCDI III.

\newpage
\begin{figure}
\begin{center}
\includegraphics*[scale=0.6,angle=-90]{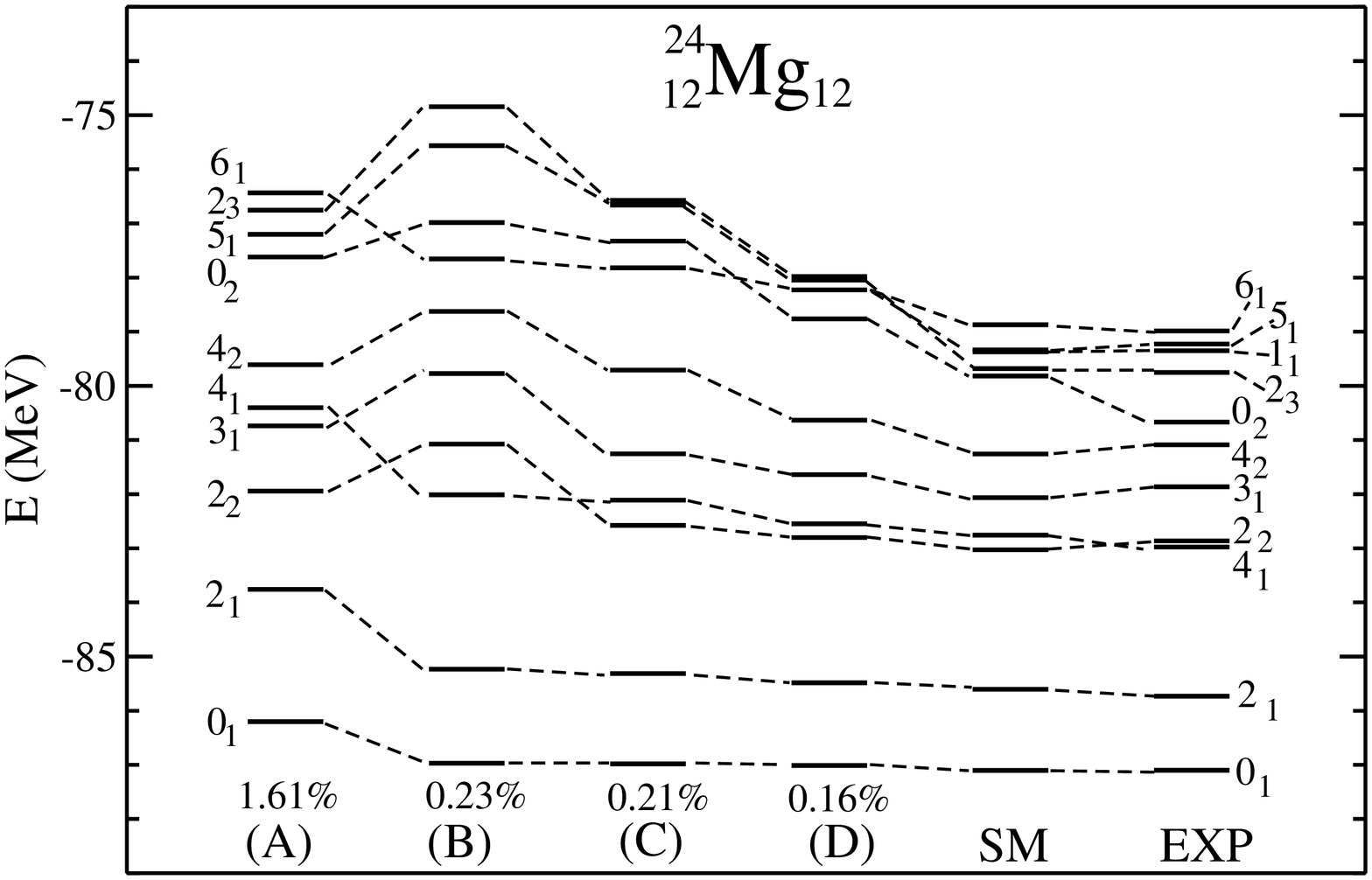}
\caption{Spectra of $^{24}$Mg obtained by performing configuration-interaction calculations in spaces
built with various sets of $T=0$ quartets (see text): (A), $J=0,2,4$ static quartets
from $^{20}$Ne; (B), $J=0,2,4$ dynamical quartets from the ground intrinsic state; (C), the same set as in 
(B) plus $J=2,3,4$ quartets from the $\gamma$ intrinsic state; (D), the same sets as in (C) plus 
$J=0,2,4$ quartets from the $\beta$ intrinsic state. SM, shell model results; EXP, experimental spectrum.
The numbers above the symbols (A)-(D) are the relative errors in the ground state correlation energy
with respect to the shell model value.}
\end{center}
\end{figure}

\newpage
\begin{figure}
\begin{center}
\includegraphics*[scale=0.6,angle=-90]{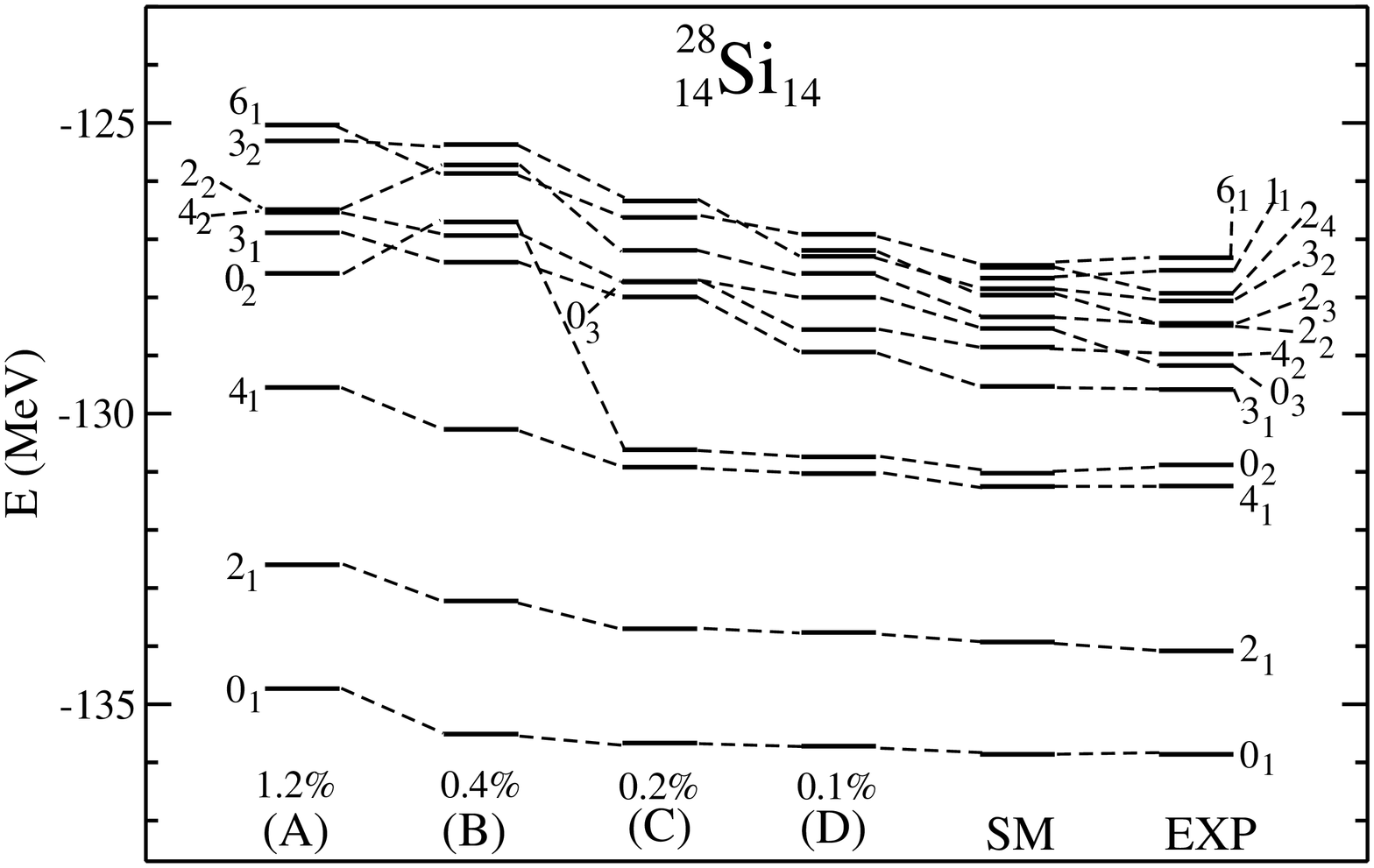}
\caption{Spectra of $^{28}$Si obtained by performing configuration-interaction calculations in spaces
built with various sets of $T=0$ quartets (see text): (A), $J=0,2,4$ static quartets
from $^{20}$Ne; (B), $J=0,2,4$ dynamical quartets from the ground intrinsic state; (C), the same set as in 
(B) plus $J=0,2,4$ quartets from the $\beta$ intrinsic state; (D), the same sets as in (C) plus 
$J=3,4$ quartets from the $k=3$ intrinsic state. SM, shell model results; EXP, experimental spectrum.
The numbers above the symbols (A)-(D) are the relative errors in the ground state correlation energy
with respect to the shell model value.}
\end{center}
\end{figure}

\newpage
\begin{figure}
\begin{center}
\includegraphics*[scale=0.6,angle=-90]{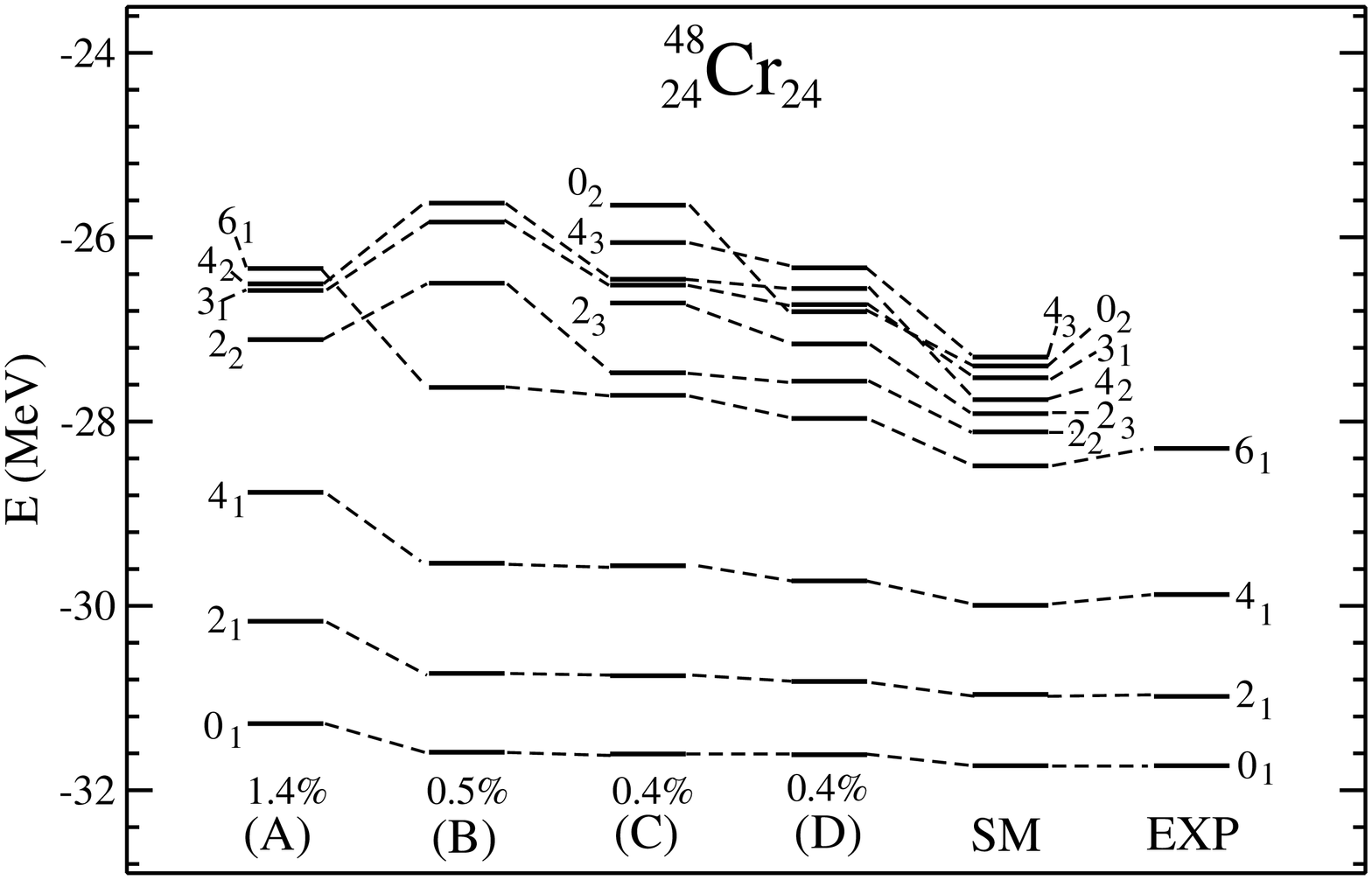}
\caption{Spectra of $^{48}$Cr obtained by performing configuration-interaction calculations in spaces
built with various sets of $T=0$ quartets (see text): (A), $J=0,2,4,6$ static quartets
from $^{44}$Ti; (B), $J=0,2,4,6$ dynamical quartets from the ground intrinsic state; (C), the same set as in 
(B) plus $J=2,3,4$ quartets from the $\gamma$ intrinsic state; (D), the same sets as in (C) plus 
$J=0,2,4$ quartets from the $\beta$ intrinsic state. SM, shell model results; EXP, experimental spectrum.
The numbers above the symbols (A)-(D) are the relative errors in the ground state correlation energy
with respect to the shell model value.}
\end{center}
\end{figure}

\newpage
\begin{figure}
\begin{center}
\includegraphics*[scale=0.6,angle=-90]{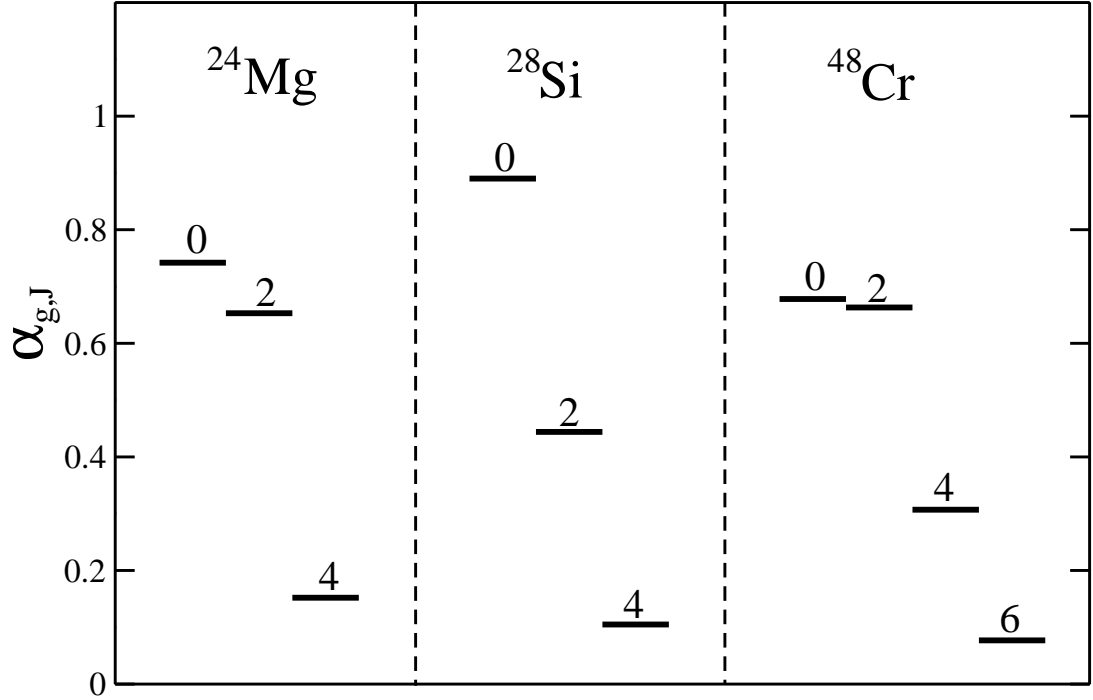}
\caption{Amplitudes $\alpha_{g,J}$ characterizing the quartets $Q^\dag_g$ of the ground intrinsic state (\ref{2})
in the three nuclei under study.}
\end{center}
\end{figure}

\end{document}